\documentclass[aos,MSNbibl,nameyear,rotating,dvips]{arximspdf}
\usepackage{mathrsfs}
\usepackage{graphicx}

\doi{10.1214/13-AOS1131} 
\volume{41}
\issue{3}
\pubyear{2013}
\firstpage{1569}
\lastpage{1592}

\makeatletter
\renewcommand{\citep}[1]{(\citeauthor{#1} \citeyear{#1})}
\newtheorem{theorem}{Theorem}[section]

\newtheorem{lemma}[theorem]{Lemma}

\newtheorem{algorithm}[theorem]{Algorithm}
\newproclaim{definition}{Definition}
\newproclaim{notation}{Notation}
\newproclaim{remark}{Remark}
\newproclaim{example}[theorem]{Example}
\newproclaim{examples}[theorem]{Examples}

\newcommand{\Var}{\operatorname{Var}}

\newcommand{\ds}{ }
\renewcommand{\u}{\cup}
\newcommand{\x}{\times}
\renewcommand{\t}{\tau}
\newcommand{\sbs}{\subset}
\renewcommand{\O}{\mathrm{O}}
\newcommand{\transpose}{\mathrm{T}}

\newcommand{\D}{\mathcal{D}}

\newcommand{\N}{\mathbb{N}}
\newcommand{\Z}{\mathbb{Z}}
\newcommand{\PP}{\mathbb{P}}
\renewcommand{\P}{\PP}
\newcommand{\E}{\mathbb{E}}

\newcommand{\s}{\mathbf{s}}
\renewcommand{\u}{\mathbf{u}}
\renewcommand{\v}{\mathbf{v}}
\newcommand{\A}{\mathbf{A}}
\newcommand{\boldS}{\mathbf{S}}
\newcommand{\0}{\mathbf{0}}
\newcommand{\1}{\mathbf{1}}
\newcommand{\bq}{\bar\mathbf{q}}
\newcommand{\br}{\bar\mathbf{r}}
\newcommand{\bv}{{\bar\mathbf{v}}}
\newcommand{\bN}{\bar N}
\newcommand{\bM}{\bar M}
\newcommand{\reduce}{\setminus}
\makeatother

\begin{document}
\begin{frontmatter}

\title{Exact sampling and counting for fixed-margin matrices}
\runtitle{Fixed-margin matrices}

\begin{aug}
\author{\fnms{Jeffrey W.} \snm{Miller}\corref{}\ead[label=e1]{Jeffrey\_Miller@Brown.edu}\thanksref{t1,t2}}
\and
\author{\fnms{Matthew T.} \snm{Harrison}\thanksref{t1}\ead[label=e2]{Matthew\_Harrison@Brown.edu}}
\thankstext{t1}{Supported in part by NSF Grant DMS-10-07593 and DARPA contract FA8650-11-1-715.}
\thankstext{t2}{Supported by a NDSEG fellowship.}
\runauthor{J. W. Miller and M. T. Harrison}
\affiliation{Brown University}
\address{Division of Applied Mathematics\\
Brown University\\
Providence, Rhode Island 02912\\
USA\\
\printead{e1}\\
\phantom{E-mail:\ }\printead*{e2}}
\end{aug}

\received{\smonth{1} \syear{2013}}
\revised{\smonth{5} \syear{2013}}

%
\begin{abstract}
The uniform distribution on matrices with specified row and column sums
is often a natural choice of null model when testing for structure in
two-way tables (binary or nonnegative integer). Due to the difficulty
of sampling from this distribution, many approximate methods have been
developed. We will show that by exploiting certain symmetries, exact
sampling and counting is in fact possible in many nontrivial real-world
cases. We illustrate with real datasets including ecological
co-occurrence matrices and contingency tables.
\end{abstract}

%


\end{frontmatter}

\section{A motivating example}
\label{sectionmotivating-example}

In ecology, co-occurrence tables are used to summarize biogeographical
data. For instance, {Table~\ref{tablemammals}} indicates the
presence/absence of 26 mammalian species in 28 mountain ranges in the
American Southwest. When presented with such data, one might wonder:
What factors control which species live in which habitats? In 1975,
ecologist (and now, renowned author) Jared Diamond stunned the ecology
community with the proposal of specific ``assembly rules'' governing
the allocation of species to habitats. \citet{Diamond1975} observed
that certain pairs of species tended to occur together, and other pairs
tended to be disjoint, suggesting that cooperation and competition play
a key role. But did these patterns really reflect species interactions,
or were they merely due to random chance?

%
\begin{sidewaystable}
\tabcolsep=0pt
\tablewidth=\textwidth
\caption{26 mammalian species in 28 mountain ranges [Patterson and Atmar
(\citeyear{Patterson1986})]}\label{tablemammals}
\begin{tabular*}{\textwidth}{@{\extracolsep{\fill
}}lcccccccccccccccccccccccccccc@{}}
\hline
& \multicolumn{28}{c@{}}{\textbf{Habitat}} \\[-6pt]
& \multicolumn{28}{c@{}}{\hrulefill} \\
\textbf{Species} & \textbf{01} & \textbf{02} & \textbf{03} &
\textbf{04} & \textbf{05} & \textbf{06} & \textbf{07} & \textbf
{08} &
\textbf{09} & \textbf{10} & \textbf{11} & \textbf{12} & \textbf
{13} & \textbf{14} & \textbf{15} & \textbf{16} & \textbf{17} &
\textbf{18} & \textbf{19} &
\textbf{20} & \textbf{21} & \textbf{22} &
\textbf{23} & \textbf{24} & \textbf{25} & \textbf{26} & \textbf
{27} & \textbf{28} \\
\hline
A & 1 & 1 & 1 & 1 & 1 & 1 & 1 & 1 & 1 & 1 & 1 & 1 & 1 & 1 & 1 & 1 & 1 &
1 & 1 & 1 & 1 & 1 & 1 & 1 & 1 & 1 & 0 & 0 \\
B & 1 & 1 & 1 & 1 & 1 & 1 & 1 & 1 & 1 & 1 & 1 & 1 & 1 & 1 & 1 & 1 & 1 &
1 & 1 & 1 & 1 & 1 & 1 & 1 & 1 & 1 & 0 & 0 \\
C & 1 & 1 & 1 & 1 & 1 & 1 & 1 & 1 & 1 & 1 & 1 & 1 & 1 & 1 & 1 & 1 & 1 &
1 & 1 & 1 & 1 & 1 & 1 & 0 & 0 & 0 & 1 & 1 \\
D & 1 & 1 & 1 & 1 & 1 & 1 & 1 & 1 & 1 & 1 & 1 & 1 & 1 & 0 & 1 & 1 & 1 &
1 & 1 & 1 & 1 & 1 & 0 & 1 & 0 & 0 & 0 & 0 \\
E & 1 & 1 & 1 & 1 & 1 & 1 & 1 & 1 & 1 & 1 & 1 & 1 & 1 & 1 & 1 & 1 & 1 &
1 & 1 & 1 & 0 & 1 & 0 & 1 & 0 & 0 & 0 & 0 \\
F & 1 & 1 & 1 & 1 & 1 & 0 & 0 & 0 & 1 & 1 & 1 & 1 & 1 & 0 & 1 & 1 & 0 &
1 & 1 & 1 & 1 & 0 & 1 & 0 & 1 & 0 & 0 & 0 \\
G & 1 & 1 & 1 & 1 & 1 & 1 & 1 & 0 & 1 & 1 & 1 & 1 & 0 & 0 & 0 & 0 & 1 &
0 & 0 & 0 & 0 & 0 & 0 & 0 & 0 & 0 & 0 & 0 \\
H & 1 & 1 & 1 & 1 & 1 & 1 & 0 & 1 & 1 & 0 & 1 & 0 & 1 & 0 & 0 & 0 & 0 &
1 & 1 & 0 & 0 & 0 & 0 & 0 & 0 & 0 & 0 & 0 \\
I & 1 & 1 & 1 & 1 & 1 & 1 & 1 & 1 & 0 & 1 & 1 & 0 & 0 & 1 & 1 & 0 & 0 &
0 & 0 & 0 & 0 & 0 & 0 & 0 & 0 & 0 & 0 & 0 \\
J & 1 & 1 & 1 & 1 & 1 & 1 & 1 & 1 & 0 & 1 & 1 & 0 & 0 & 1 & 0 & 0 & 0 &
0 & 0 & 0 & 0 & 0 & 0 & 0 & 0 & 0 & 0 & 0 \\
K & 1 & 1 & 1 & 1 & 1 & 1 & 1 & 1 & 1 & 1 & 0 & 0 & 0 & 0 & 0 & 0 & 0 &
0 & 0 & 0 & 0 & 0 & 0 & 0 & 0 & 0 & 0 & 0 \\
L & 1 & 1 & 1 & 1 & 1 & 0 & 0 & 1 & 1 & 0 & 0 & 0 & 1 & 0 & 0 & 0 & 0 &
0 & 0 & 0 & 1 & 0 & 1 & 0 & 0 & 0 & 0 & 0 \\
M & 1 & 1 & 1 & 1 & 1 & 0 & 0 & 0 & 0 & 0 & 0 & 1 & 0 & 0 & 0 & 1 & 1 &
0 & 0 & 0 & 0 & 0 & 0 & 0 & 0 & 0 & 0 & 0 \\
N & 1 & 1 & 1 & 1 & 0 & 1 & 1 & 1 & 0 & 0 & 0 & 0 & 0 & 1 & 0 & 0 & 0 &
0 & 0 & 0 & 0 & 0 & 0 & 0 & 0 & 0 & 0 & 0 \\
O & 1 & 1 & 1 & 1 & 1 & 1 & 1 & 1 & 0 & 0 & 0 & 0 & 0 & 0 & 0 & 0 & 0 &
0 & 0 & 0 & 0 & 0 & 0 & 0 & 0 & 0 & 0 & 0 \\
P & 1 & 1 & 0 & 1 & 1 & 1 & 0 & 0 & 1 & 0 & 0 & 0 & 1 & 0 & 0 & 0 & 0 &
0 & 0 & 0 & 0 & 0 & 0 & 0 & 0 & 0 & 0 & 0 \\
Q & 1 & 1 & 1 & 1 & 1 & 0 & 1 & 0 & 0 & 0 & 0 & 0 & 0 & 0 & 0 & 0 & 0 &
0 & 0 & 0 & 0 & 0 & 0 & 0 & 0 & 0 & 0 & 0 \\
R & 1 & 1 & 1 & 1 & 1 & 0 & 0 & 0 & 0 & 0 & 0 & 1 & 0 & 0 & 0 & 0 & 0 &
0 & 0 & 0 & 0 & 0 & 0 & 0 & 0 & 0 & 0 & 0 \\
S & 1 & 1 & 1 & 1 & 1 & 0 & 0 & 0 & 0 & 0 & 0 & 0 & 0 & 0 & 0 & 0 & 0 &
0 & 0 & 0 & 0 & 0 & 0 & 0 & 0 & 0 & 0 & 0 \\
T & 1 & 1 & 1 & 1 & 1 & 0 & 0 & 0 & 0 & 0 & 0 & 0 & 0 & 0 & 0 & 0 & 0 &
0 & 0 & 0 & 0 & 0 & 0 & 0 & 0 & 0 & 0 & 0 \\
U & 1 & 1 & 1 & 1 & 0 & 0 & 0 & 0 & 0 & 0 & 0 & 0 & 0 & 0 & 0 & 0 & 0 &
0 & 0 & 0 & 0 & 0 & 0 & 0 & 0 & 0 & 0 & 0 \\
V & 1 & 1 & 1 & 0 & 0 & 0 & 1 & 0 & 0 & 0 & 0 & 0 & 0 & 0 & 0 & 0 & 0 &
0 & 0 & 0 & 0 & 0 & 0 & 0 & 0 & 0 & 0 & 0 \\
W & 1 & 1 & 1 & 0 & 0 & 0 & 0 & 0 & 0 & 0 & 0 & 0 & 0 & 0 & 0 & 0 & 0 &
0 & 0 & 0 & 0 & 0 & 0 & 0 & 0 & 0 & 0 & 0 \\
X & 1 & 1 & 1 & 0 & 0 & 0 & 0 & 0 & 0 & 0 & 0 & 0 & 0 & 0 & 0 & 0 & 0 &
0 & 0 & 0 & 0 & 0 & 0 & 0 & 0 & 0 & 0 & 0 \\
Y & 1 & 0 & 0 & 0 & 0 & 0 & 0 & 0 & 0 & 0 & 0 & 0 & 0 & 0 & 0 & 0 & 0 &
0 & 0 & 0 & 0 & 0 & 0 & 0 & 0 & 0 & 0 & 0 \\
Z & 1 & 0 & 0 & 0 & 0 & 0 & 0 & 0 & 0 & 0 & 0 & 0 & 0 & 0 & 0 & 0 & 0 &
0 & 0 & 0 & 0 & 0 & 0 & 0 & 0 & 0 & 0 & 0 \\
\hline
\end{tabular*}
\end{sidewaystable}

To address this question, \citet{Connor1979} suggested statistical
hypothesis testing. Since some species are simply more prolific than
others, and some habitats are larger than others, a sensible choice of
null model is the uniform distribution on co-occurrence matrices with
the observed numbers of habitats per species (row sums) and species per
habitat (column sums). \citet{Connor1979} presented a formidable
challenge to Diamond's theory by showing that under this simple null
model, the statistics observed by Diamond could easily have arisen by
chance. A~contentious debate erupted, yielding an extensive body of
research on test statistics and null models for detecting various types
of ``structure'' in ecological matrices. Decades later, the basic null
model of Connor and Simberloff has withstood the test of time, and is
now a mainstay in the analysis of ecological matrices; see, for
example, \citet{Ulrich2007,Gotelli2002} and references therein.

As a concrete example, consider the montane mammals in {Table~\ref{tablemammals}}. \citet{Patterson1986} proposed a model in
which, during the most recent glacial period, cold-adapted species
inhabited a region spanning several mountain ranges and the low-lying
areas between, but in the current (warmer) interglacial period these
populations have receded into the mountains and become extinct in some
areas. They suggest that this would cause the set of species found in
one mountain range to tend to be a subset of those found in another.
This led them to consider the following \textit{nested subset statistic},
equal to the number of species--habitat pairs such that the species does
not occur in that habitat but does occur in a less-populated habitat,
\[
S_{\mathrm{nest}} = \sum_{i,j} I(a_{ij} =
0, q_j>m_i),
\]
where $\A=(a_{ij})$ is a binary matrix with species as rows and
habitats as columns (such as {Table~\ref{tablemammals}}),
$q_j=\sum_i a_{ij}$ and $m_i =\min\{q_j\dvtx a_{ij}=1, j= 1,\ldots,n\}$.
Here, $I(E)$ is $1$ if $E$ is true and $0$ otherwise. (Note: Smaller
$S_{\mathrm{nest}}$ means more ``nestedness.'')
To perform a hypothesis test using the standard null model described
above, one would estimate the $p$-value for $S_{\mathrm{nest}}$ by
sampling from
the uniform distribution over binary matrices with the observed row and
column sums---in this case, (26, 26, 25, 22, 22, 18, 12, 12, 12, 11,
10, 10, 8, 8, 8, 7, 6, 6, 5, 5, 4, 4, 3, 3, 1, 1) and (26, 24, 23, 21,
19, 13, 13, 12, 11, 10, 10, 9, 9, 7, 7, 7, 7, 7, 7, 6, 6, 5, 5, 4, 3,
2, 1, 1), respectively. However, it is difficult to sample exactly from
this distribution. Instead, Patterson and Atmar used an approximation
in which the entries of each column are drawn proportionally to the row
sums, conditioned on the column sum. (The row sums are not constrained
in their approximation.) They drew 1000 samples from their
approximation, estimated the $p$-value of $S_{\mathrm{nest}}$ to be
$9\times
10^{-20}$ for {Table~\ref{tablemammals}}, and concluded that the
data does exhibit significantly more nestedness than one would expect
under the null.
Patterson and Atmar's (\citeyear{Patterson1986}) article was highly
influential, inspiring many subsequent studies into nestedness; see,
for example, \citet{Ulrich2007} and references therein.

The preceding scenario is commonplace---the combinatorial problem
that arises from constraining the row and column sums makes it
difficult to sample exactly from\vadjust{\goodbreak} the desired uniform distribution. As a
result, on all but the most trivial matrices, researchers have resorted
to approximate methods, such as Markov chain Monte Carlo (MCMC),
sequential importance sampling and heuristic approaches such as the
one described above. With all these approximate methods, the nagging
question remains: Did the use of an approximate distribution
significantly affect the result?

In this work, we describe an efficient algorithm for sampling exactly
from the uniform distribution over binary or nonnegative integer
matrices with given row and column sums (provided that most of the sums
are not too large). As a result, Monte Carlo estimates of quantities of
interest, such as $p$-values, can be accompanied by exact confidence
intervals (i.e., true confidence intervals, rather than intervals based
on asymptotic approximations). Further, our algorithm computes the
exact number of such matrices.

%
\begin{table}
\caption{Sample statistics of $S_{\mathrm{nest}}$ for montane mammal data
({Table~\protect\ref{tablemammals}})}\label{tablemammal-results}
\begin{tabular*}{\textwidth}{@{\extracolsep{\fill}}lcccccc@{}}
\hline
\textbf{Method} & \textbf{\# samples} & \textbf{estimated $p$-value}
& \textbf{mean} & \textbf{std. dev.} & \textbf{min} &
\textbf{max}\\
\hline
Exact & 1,000,000 & $0.0322 \pm0.00018$ & \phantom{0}80.7 & \phantom
{0}9.7 & \phantom{0}44 & 132 \\
Heuristic & 1000 & $9\times10^{-20}$ & 227.9 & 18.1 & 180 & 287 \\
\hline
\end{tabular*}
\end{table}

\begin{figure}

\includegraphics{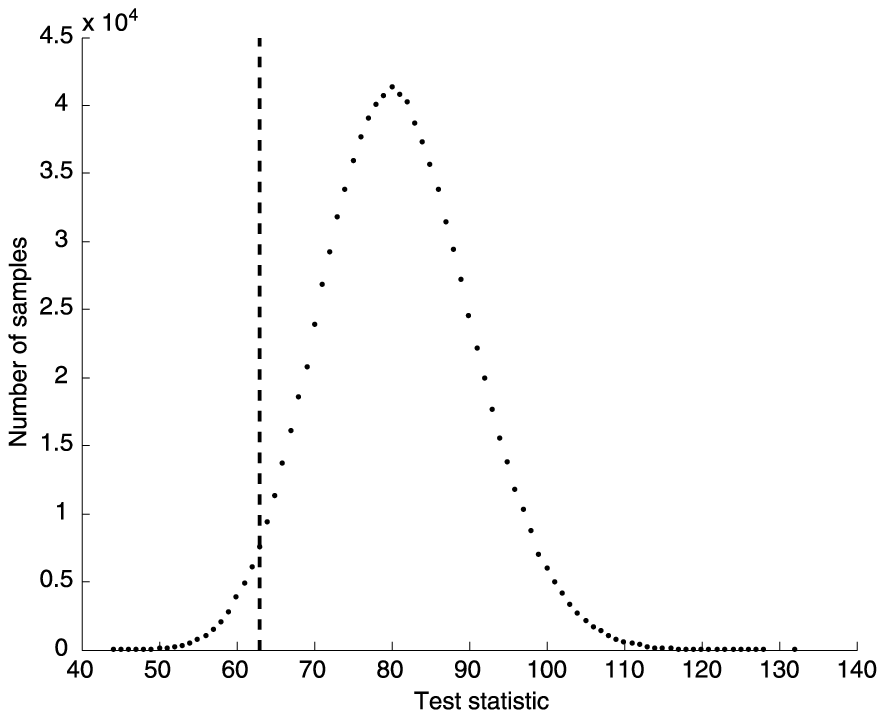}

\caption{Histogram of $S_{\mathrm{nest}}$ for $10^6$ exact samples
from the
uniform distribution over matrices with
margins as in {Table~\protect\ref{tablemammals}}. The dashed line
is the value of $S_{\mathrm{nest}}$ for {Table~\protect\ref{tablemammals}}.}
\label{figuremammal-results}
\end{figure}

In {Table~\ref{tablemammal-results}}, we compare the results of
Patterson and Atmar with results based on $10^6$ exact samples using
our algorithm. There are large discrepancies. We estimate the $p$-value
to be $0.0322 \pm0.00018$ ($\hat p\pm\hat{\mathrm{se}}$), with an exact $95\%$
confidence interval of $[0.0318, 0.0326]$ (based on the binomial
c.d.f., not on $\hat{\mathrm{se}}$). Their estimate of $9\times10^{-20}$ is far
smaller---in fact, the value of the test statistic on the observed
matrix, $S_{\mathrm{nest}} = 63$, is considerably less than the
smallest value
among all of their 1000 samples, $S_{\mathrm{nest}} = 180$. (Note: It appears
that they used a normal approximation to the distribution of
$S_{\mathrm{nest}}$
to estimate the $p$-value.) Meanwhile, in view of the histogram of
exact samples in {Figure~\ref{figuremammal-results}}, the
observed matrix appears relatively typical! Recall that in their
approximation, the entries of each column are drawn proportionally to
the row sums, conditioned on the column sum, and that the row sums are
not constrained. Apparently, omitting the constraint on the row sums
has a drastic effect. As a result, the analysis dramatically
underestimated the $p$-value. This vividly illustrates the utility of
exact sampling in these problems.

Our algorithm required 46 seconds to find that there are
$2\mbox{,}663\mbox{,}296\mbox{,}694\mbox{,}330\mbox{,}\break 271\mbox{,}332\mbox{,}856\mbox{,}672\mbox{,}902\mbox{,}543\mbox{,}209\mbox{,}853\mbox{,}700$
($\approx\!2.7\times10^{39}$) binary matrices with row and column sums
as in {Table~\ref{tablemammals}}, and subsequently, required 4.2
milliseconds per exact sample. (All computations reported in this paper
were performed using a 2.8~GHz processor with 6 GB of RAM.)
We know of no other algorithm capable of exact counting and sampling
for matrices of this size.

%

\section{Overview}
\label{sectionoverview}

Let $N(\mathbf{p},\mathbf{q})$ be the number of $m\x n$ binary
matrices with margins
(row and column sums) $\mathbf{p}=(p_1,\ldots,p_m)\in\N^m$ and
$\mathbf{q}
=(q_1,\ldots
,q_n)\in\N^n$, respectively, and let $M(\mathbf{p},\mathbf{q})$ be
the corresponding
number of $\N$-valued matrices. (We use $\N=\{0,1,\ldots\}$ throughout.)
In this paper we develop a technique for finding $N(\mathbf{p},\mathbf{q})$
and $M(\mathbf{p},\mathbf{q}
)$, and for exact uniform sampling from these sets of matrices---an
important and challenging problem
[\citet{Diaconis1995}, \citet{Chen2005}].
Our method is feasible for modestly-sized matrices (roughly, $m + n\leq
100$ with current desktop computing power) or very sparse large matrices.

As described above, in the binary case, an important application is
testing for structure in ecological co-occurrence matrices. It turns
out that most real-world ecology matrices are small enough that our
method is feasible. In the $\N$-valued case, an important application
is the conditional volume test of \citet{Diaconis1985} for two-way
contingency tables (for which our method is feasible as long as the
margins are relatively small). In addition to these direct
applications, a major auxiliary benefit of having an exact method is
that it enables one to measure the accuracy of certain approximate
methods (which can scale to matrices far larger than our exact method
can accommodate).

Since a bipartite graph with degree sequences $\mathbf{p}=(p_1,\ldots
,p_m)\in\N
^m$, $\mathbf{q}=(q_1,\ldots,q_n)\in\N^n$
(and $m,n$ vertices in each part, resp.) can be viewed as a $m\times n$
matrix with row and column sums $(\mathbf{p},\mathbf{q})$, our
technique applies
equally well to counting and uniformly sampling such bipartite graphs.\vadjust{\goodbreak}
Under this correspondence, simple graphs correspond to binary matrices,
and multigraphs correspond to $\N$-valued matrices.

The distinguishing characteristic of our method is its tractability on
matrices of nontrivial size.
In general, computing $M(\mathbf{p},\mathbf{q})$ is \#P-complete
[\citet{Dyer1997}],
and perhaps $N(\mathbf{p},\mathbf{q})$ is as well.
However, if one assumes a bound on the column sums
then our algorithm computes both numbers in polynomial time.
After counting, uniform samples may be drawn in polynomial expected
time for bounded column sums.
To our knowledge, all previous algorithms \textit{for the nonregular case}
require super-polynomial time (in the worst case) to compute these
numbers, even for bounded column sums.
(We assume a description length of at least $m+n$
and no more than $m\log a+n\log b$, where $a=\max p_i$, $b=\max q_i$.)
In general (without assuming a bound on the column sums),
our algorithm computes $N(\mathbf{p},\mathbf{q})$ or $M(\mathbf
{p},\mathbf{q})$ in
$\O
(m(ab+c)(a+b)^{b-1}(b+c)^{b-1}(\log c)^3)$ time
for $m\x n$ matrices,
where $a=\max p_i$, $b=\max q_i$ and $c=\sum p_i=\sum q_i$.
After counting, uniform samples may be drawn in $\O(mc\log c)$
expected time.

In complement to most approaches to exactly computing $M(\mathbf
{p},\mathbf{q}
)$, which
are efficient for small tables with large margins, our algorithm is
efficient for large tables with small margins.
For instance, computing $M(\mathbf{p},\mathbf{q})$ for the $100\times100$
matrices with
$\mathbf{p}=\mathbf{q}=(5^{(20)}, 4^{(20)}, 3^{(20)}, 2^{(20)}, 1^{(20)})$,
where $x^{(k)}$ denotes $x$ repeated $k$ times, takes 701 seconds (the
exact number, approximately $2.9580567\times10^{434}$, is available
upon request). Likewise, computing $N(\mathbf{p},\mathbf{q})$ for the same
$\mathbf{p}$ and $\mathbf{q}
$ takes 688 seconds (the number is approximately $2.3514766\times10^{431}$).

The remainder of the paper is organized as follows. In the rest of this
section, we describe related work from the literature. In {Section~\ref{sectionresults}}, we first describe the intuition behind our
technique, then formally state the recursions and the resulting
algorithm, and give bounds on computation time. In {Section~\ref{sectionapplications}}, we illustrate some applications. In
{Section~\ref{sectionproofs}} we prove the recursions and in
{Section~\ref{sectioncomputation}}, we prove the bounds on
computation time.

\subsection{Previous work}

We briefly survey the previous work on this problem.
This review is not exhaustive, focusing instead on those results which
are particularly significant
or closely related to the present work.
Let $H_n(r)$ and $H^*_n(r)$ denote $M(\mathbf{p},\mathbf{q})$ and
$N(\mathbf{p},\mathbf{q}
)$, respectively, when
$\mathbf{p}=\mathbf{q}=(r,\ldots,r)\in\N^n$.
The predominant focus has been on the regular cases $H_n(r)$ and $H^*_n(r)$.

Work on counting these matrices goes back at least as far as \citeauthor{MacMahon1915}
[(\citeyear{MacMahon1915}), see Volume II, page 161], who applied his expansive
theory to find the polynomial for $H_3(r)$.
Redfield's theorem [\citet{Redfield1927}], inspired by MacMahon,
can be used to derive summations for some special cases,
such as $H_n(r), H^*_n(r)$ for $r=2,3$, and in similar work, Read
(\citeyear
{Read1959,Read1960}) used P\'{o}lya theory to derive these summations
for $r=3$.
Two beautiful theoretical results must also be mentioned:
\citet{Stanley1973} proved that for fixed $n$, $H_n(r)$ is a
polynomial in
$r$, and
Gessel (\citeyear{Gessel1987,Gessel1990}) showed that for fixed $r$,\vadjust{\goodbreak}
both $H_n(r)$ and $H^*_n(r)$ are $P$-recursive in $n$,
vastly generalizing the recursions for $H_n(2)$, $H^*_n(2)$ found by
\citet{Anand1966}.

We turn next to algorithmic results more closely related to the present work.
\citet{McKay1983} and \citet{Canfield2005} have demonstrated a
coefficient extraction technique
for computing $N(\mathbf{p},\mathbf{q})$ in the semi-regular case [in which
$\mathbf{p}
=(a,\ldots,a)\in\N^m$ and $\mathbf{q}=(b,\ldots,b)\in\N^n$].
To our knowledge,
McKay's is the most efficient method known previously for $N(\mathbf
{p},\mathbf{q})$.
By our analysis it requires at least $\Omega(mn^{b})$ time for bounded $a,b$,
while the method presented here is $\O(mn^{b}(\log n)^3)$ in this case.
Since this latter bound is quite crude, we expect that our method
should have comparable or better performance, and indeed empirically we
find that typically it is more efficient.
If only $b$ is bounded, McKay's algorithm is still $\Omega(mn^b)$,
but the bound on our performance increases to $\O(mn^{2b-1}(\log n)^3)$,
so it is possible that McKay's algorithm will outperform ours in these cases.
Nonetheless, it is important to bear in mind that McKay's algorithm
is efficient only in the semi-regular case (while our method permits
nonregular margins).
If neither $a$ nor $b$ is bounded, McKay's method is exponential in $b$
(as is ours).

\citet{mckay1990uniform} presented an intriguing randomized algorithm
for exact uniform sampling from the set of binary matrices with margins
$\mathbf{p},\mathbf{q}$, provided that all of the margins are sufficiently
small. It
takes $\O((m+n)^2 k^4)$ expected time per sample, where $k$ is an upper
bound on all of the margins. Unfortunately, the conditions under which
it applies are rather restrictive---in fact, for most of the
real-world problems we have encountered, it reduces to a simple
rejection sampling scheme that is rather inefficient.

Regarding $M(\mathbf{p},\mathbf{q})$, one of the most efficient algorithms
known to
date is LattE (lattice point enumeration) [\citet{Deloera2004}], which
uses the algorithm of \citet{Barvinok1994} to count lattice points
contained in convex polyhedra. It runs in polynomial time for any fixed
dimension, and as a result it can compute $M(\mathbf{p},\mathbf{q})$
for astoundingly
large margins, provided that $m$ and $n$ are small. However, since the
computation time grows very quickly with the dimension, LattE is
currently inapplicable when $m$ and $n$ are larger than $6$.
There are similar algorithms [\citet{Mount2000}, \citet{Deloera2003},
\citet{Beck2003}] that are efficient for small matrices.

In addition, several other algorithms have been presented for finding
$N(\mathbf{p},\mathbf{q})$ [such as \citet
{Johnsen1987,Wang1988,Wang1998,Perez2002}] 
and $M(\mathbf{p},\mathbf{q})$ [\citet{Gail1977,Diaconis1995}] allowing
nonregular margins;
however, it appears that all are exponential in the size of the matrix,
even for bounded margins.
While in this work we are concerned solely with exact results,
we note that many useful approximations
for $N(\mathbf{p},\mathbf{q})$ and $M(\mathbf{p},\mathbf{q})$ (in
the general case)
have been found,
as well as approximate sampling algorithms [\citet
{Holmes1996,Chen2005,Greenhill2006,Canfield2008,Harrison2013}].

\section{Main results}
\label{sectionresults}

\subsection{Idea of the algorithm}
\label{sectionidea}

Before formally presenting the results, we introduce the simple
observation underlying our approach. The idea of recursively counting
these matrices is old [\citet{Gail1977}]; the key novelty of our
method is that it exploits symmetries arising from repeated column
sums, leading to a dramatic improvement over na\"ive recursions.
The basic idea is very straightforward---the surprising part is that
it leads to an algorithm that is efficient for many nontrivial datasets.

For simplicity, consider the binary case. The first thing to notice is
that $N(\mathbf{p},\mathbf{q})$ is unchanged by permutations of the entries
of $\mathbf{p}$ or
$\mathbf{q}$ (as is easy to see).
Now, suppose $\mathbf{p}=(6,6,4,3,1,1,1)$ and $\mathbf{q}
=(0,0,1,1,1,1,2,2,2,3,3,3,3)$, and consider the partially-filled matrix
in {Table~\ref{tableidea}} in which the first row is $\u
=(0,0,1,1,0,1,0,1,0,1,0,1,0)$ and the rest of the matrix is undetermined.


\begin{table}[t]
\caption{Partially-filled matrix}\label{tableidea}
\begin{tabular}{|c|c||c|c|c|c||c|c|c||c|c|c|c|c}
\cline{1-13}
0 & 0 & 1 & 1 & 0 & 1 & 0 & 1 & 0 & 1 & 0 & 1 & 0 &  6 \\ \cline{1-13}
  &   &   &   &   &   &   &   &   &   &   &   &   &  6 \\
  &   &   &   &   &   &   &   &   &   &   &   &   &  4 \\
  &   &   &   &   &   &   &   &   &   &   &   &   &  3 \\
  &   &   &   &   &   &   &   &   &   &   &   &   &  1 \\
  &   &   &   &   &   &   &   &   &   &   &   &   &  1 \\
  &   &   &   &   &   &   &   &   &   &   &   &   &  1 \\
\cline{1-13}
\multicolumn{1}{c}{0} &
\multicolumn{1}{c||}{0} &
\multicolumn{1}{c}{1} &
\multicolumn{1}{c}{1} &
\multicolumn{1}{c}{1} &
\multicolumn{1}{c||}{1} &
\multicolumn{1}{c}{2} &
\multicolumn{1}{c}{2} &
\multicolumn{1}{c||}{2} &
\multicolumn{1}{c}{3} &
\multicolumn{1}{c}{3} &
\multicolumn{1}{c}{3} &
\multicolumn{1}{c}{3} &
\end{tabular}
\label{table:idea}
\end{table}

There are $N(\mathbf{p},\mathbf{q})$ binary matrices with margins
$(\mathbf
{p},\mathbf{q})$, and there
are clearly $N(L\mathbf{p},\mathbf{q}-\u)$ of these with $\u$ as
the first
row, where
$L$ denotes the left-shift map: $L\mathbf{p}=(p_2,\ldots,p_m)$.
Divide the
first row into blocks $0,1,2,\ldots,m$ where block $k$ contains the
columns $i$ such that $q_i = k$. Since the number of matrices is
unchanged under permutations of the margins, then $N(L\mathbf
{p},\mathbf{q}
-\u)=N(L\mathbf{p},\mathbf{q}
-\v)$ for any permutation $\v$ of $\u$ such that the number of ones in
each block is unchanged. If $r_k$ is the size of block $k$, and $s_k$
is the number of ones in block $k$, then there are
\[
\pmatrix{\mathbf{r}
\cr
\s}:= \pmatrix{r_1
\cr
s_1}
\cdots\pmatrix{r_m
\cr
s_m}
\]
such permutations $\v$, where $\mathbf{r}=(r_1,\ldots,r_m)$ and $\s
=(s_1,\ldots,s_m)$. For instance, in this example, ${\mathbf
{r}\choose\s} =
{4\choose3}{3\choose1}{4\choose2}$. Note that $s_0$ will be $0$, and
thus ${r_0\choose s_0}$ will be $1$, for any $\u$ such that
$N(L\mathbf{p},\mathbf{q}-\u
)\neq0$. Therefore,
\[
N(\mathbf{p},\mathbf{q}) =\sum_{\u\in\{0,1\}^n} N(L\mathbf {p},
\mathbf{q}-\u) = \sum_\s\pmatrix{\mathbf{r}
\cr
\s}
N(L\mathbf{p},\mathbf{q}- \u_\s),
\]
where the second sum is over nonnegative integer vectors $\s
=(s_1,\ldots
,s_m)$ summing to $p_1$, and $\u_\s\in\{0,1\}^n$ is any binary vector
with $s_k$ ones in block $k$ for each $k= 1,\ldots,m$, and zero ones in
block $0$. This defines a recursion for $N(\mathbf{p},\mathbf{q})$
which, when
carefully implemented using dynamic programming and the Gale--Ryser
criterion (described below), is the basis for our algorithm. This
computation yields a data structure that enables efficient sampling in
a row-by-row fashion.
There is a similar (although more subtle) recursion for the case of
$M(\mathbf{p},\mathbf{q})$.

\subsection{Recursions, algorithms and bounds}

Introducing the following notation will be useful.
We consider $\N^n$ to be the subset of
$\N^\infty:=\{(r_1,r_2,\ldots)\dvtx r_i\in\N\mbox{ for } i =
1,2,\ldots\}$
such that all but the first $n$ components are zero.
Let $L\dvtx\N^\infty\to\N^\infty$ denote the left-shift map:
$L\mathbf{r}
=(r_2,r_3,\ldots)$.
Given $\mathbf{r},\s\in\N^n$, let
$\mathbf{r}\reduce\s:= \mathbf{r}-\s+L\s$ (which may be read as
``$\mathbf{r}$ \textit{reduce}
$\s$''),
let ${\mathbf{r}\choose\s}:={r_1\choose s_1}\cdots{r_n\choose s_n}$,
and let $\br$ denote the vector of counts,
$\br:= (\bar r_1,\bar r_2,\ldots)$ where $\bar r_i:=\#\{j\dvtx r_j=i\}$.
We write $\mathbf{r}\le\s$ if $r_i\le s_i$ for all $i$.
Given $n\in\N$, let $C_n(k):=\{\mathbf{r}\in\N^n\dvtx \sum_i
r_i=k\}$ be the
$n$-part compositions (including zeros) of $k$,
and given $\s\in\N^n$, let $C^\s(k):=\{\mathbf{r}\in C_n(k) \dvtx
\mathbf{r}\le\s\}$.
Since $N(\mathbf{p},\mathbf{q})$ and $M(\mathbf{p},\mathbf{q})$ are
fixed under permutations
of the row sums $\mathbf{p}$ or column sums $\mathbf{q}$, and since
zeros in the
margins do not affect the number of matrices, then we may define
$\bN(\mathbf{p},\bq):=N(\mathbf{p},\mathbf{q})$ and $\bM(\mathbf
{p},\bq
):=M(\mathbf{p},\mathbf{q})$ without ambiguity.

%
\begin{theorem}[(Recursions)]
\label{thmrec}
The number of matrices with margins\break $(\mathbf{p},\mathbf{q})\in\N
^m\x\N^n$
is given by
\begin{eqnarray*}
&&(1)\quad \bN(\mathbf{p},\mathbf{r})=\sum_{\s\in C^\mathbf
{r}(p_1)} \pmatrix{\mathbf{r}\cr
\s} \bN(L\mathbf{p},\mathbf{r}
\reduce\s)\quad
\mbox{for binary matrices, and}
\\
&&(2)\quad\bM(\mathbf{p},\mathbf{r})=\sum_{\s\in C^{\mathbf
{r}+L\s}(p_1)} \pmatrix{\mathbf{r}
+L\s\cr\s} \bM
(L\mathbf{p},\mathbf{r}\reduce\s)\quad
\mbox{for $\N$-valued matrices,}
\end{eqnarray*}
where $\mathbf{r}=\bq$, and in \textup{(2)}, we sum over all $\s$
such that
$\s
\in C^{\mathbf{r}+L\s}(p_1)$.
\end{theorem}

For the proof, see {Section~\ref{sectionproofs}}.
In the binary case, computation of this sum is greatly simplified by
summing only over those $\s\in C^\mathbf{r}(p_1)$ for which $\bN
(L\mathbf
{p},\mathbf{r}\reduce\s
)$ is nonzero. This can be efficiently achieved using the Gale--Ryser
criterion [\citet{Gale1957}, \citet{Ryser1957}], which provides the following
necessary and sufficient condition for the existence of a binary matrix
with margins $(\mathbf{p},\mathbf{q})$:
when $q_i':=\#\{j \dvtx q_j\geq i\}$ and $p_1\geq\cdots\geq p_m$, we have
$N(\mathbf{p},\mathbf{q})\neq0$
if and only if $\sum_{i=1}^j p_i \leq\sum_{i=1}^j q_i' \mbox{ for all
} j<m
\mbox{ and } \sum_{i=1}^m p_i = \sum_{i=1}^m q_i'$.
This is easily translated into a condition in terms of $(\mathbf
{p},\bq)$ and
$\bN(\mathbf{p},\bq)$.
In the $\N$-valued case, there is no analogue to the Gale--Ryser
criterion [since $M(\mathbf{p},\mathbf{q})>0$ for any nonnegative
$\mathbf
{p},\mathbf{q}$ such that
$\sum_i p_i =\sum_i q_i$].
The following recursive procedure can be used to compute either
$N(\mathbf{p},\mathbf{q}
)$ or $M(\mathbf{p},\mathbf{q})$.

%
\begin{algorithm}[(Counting)] \label{algc}\\
Input: $(\mathbf{p},\bq)$, where $(\mathbf{p},\mathbf{q})\in\N
^m\x\N^n$
are margins such that
$\sum_i p_i=\sum_i q_i$.\\
Output: $N(\mathbf{p},\mathbf{q})$ [or $M(\mathbf{p},\mathbf{q})$],
the number of
binary (or $\N
$-valued) matrices.\\
Storage: Lookup table initialized with $\bN(\0,\0)=1$ [or $\bM(\0,\0)=1$].
\begin{enumerate}
\item[(1)] If $\bN(\mathbf{p},\bq)$ [or $\bM(\mathbf{p},\bq)$] is in
the lookup table,
return the result.
\item[(2)] In the binary case, if Gale--Ryser gives $\bN(\mathbf{p},\bq
)=0$, store
the result and return~$0$.
\item[(3)] Evaluate the sum in {Theorem~\ref{thmrec}}, recursing to
step \textup{(1)} for each term.
\item[(4)] Store the result and return it.
\end{enumerate}
\end{algorithm}

Let $T(\mathbf{p},\mathbf{q})$ be the time (number of machine operations)
required by
{Algorithm~\ref{algc}} to compute
$N(\mathbf{p},\mathbf{q})$ or $M(\mathbf{p},\mathbf{q})$, after
performing an $\O
(n^3)$ preprocessing
step to compute all needed binomial coefficients.
[It turns out that computing $M(\mathbf{p},\mathbf{q})$ always takes longer,
but the
bounds we prove apply to both $N(\mathbf{p},\mathbf{q})$ and
$M(\mathbf
{p},\mathbf{q})$.]
We give a series of bounds on $T(\mathbf{p},\mathbf{q})$ ranging from tighter
but more
complicated,
to more crude but simpler. The bounds will absorb the $\O(n^3)$
pre-computation except in the trivial case when the maximum column sum
is $1$.

%
\begin{theorem}[(Bounds)] Suppose $(\mathbf{p},\mathbf{q})\in\N^m\x
\N^n$,
$a=\max p_i$,
$b=\max q_i$ and $c=\sum p_i =\sum q_i$. Then:
\label{thmeff}
\begin{longlist}[(1)]
\item[(1)]
$\ds T(\mathbf{p},\mathbf{q}) = \O\bigl((ab+c)(\log c)^3\sum_{i=1}^m {p_i+b-1
\choose
b-1}{p_i+\cdots+p_m + b-1 \choose b-1}\bigr)$,
\item[(2)]$\ds T(\mathbf{p},\mathbf{q}) = \O
(m(ab+c)(a+b)^{b-1}(b+c)^{b-1}(\log c)^3)$,
\item[(3)]$\ds T(\mathbf{p},\mathbf{q}) = \O(mn^{2b-1}(\log n)^3)$ for
bounded $b$,
\item[(4)]$\ds T(\mathbf{p},\mathbf{q}) = \O(mn^{b}(\log n)^3)$ for
bounded $a,b$.
\end{longlist}
\end{theorem}
For the proof, see {Section~\ref{sectioncomputation}}.
Since we may swap $\mathbf{p}$ and $\mathbf{q}$ without changing the number
of matrices,
we could use {Algorithm~\ref{algc}} on $(\mathbf{q},\bar\mathbf{p})$
to compute
$N(\mathbf{p},\mathbf{q})$ or $M(\mathbf{p},\mathbf{q})$ using
$T(\mathbf{q},\mathbf{p})$
operations, which, for
example, is $\O(nm^{a}(\log m)^3)$ for bounded $a,b$.
Typically, we find that choosing $\mathbf{p}$ to be the shorter of the two
vectors is preferable.
$T(\mathbf{p},\mathbf{q})$ also depends on the ordering of the row sums
$p_1,\ldots
,p_m$ as suggested by {Theorem~\ref{thmeff}}(1), and we find that
putting them in decreasing order $p_1\geq\cdots\geq p_m$ tends to
work well.
In the binary case, {Algorithm~\ref{algc}} is typically made
significantly more efficient
by using the Gale--Ryser conditions, and this is not accounted for in
these bounds.

It is worth mentioning that a significant further reduction in
computation time can be achieved by factoring the sums in {Theorem~\ref
{thmrec}}. For example, in the binary case, we use
\[
\bN(\mathbf{p},\mathbf{r})= \sum_{s_1}
\pmatrix{r_1
\cr
s_1} \sum_{s_2}
\pmatrix{r_2
\cr
s_2} \cdots\sum
_{s_m} \pmatrix{r_m
\cr
s_m} \bN(L
\mathbf{p},\mathbf{r}\reduce\s),
\]
where for each $k= 1,\ldots,m$, $s_k$ is summed over a range of values
chosen so that $\s$ will always satisfy both $\s\in C^\mathbf
{r}(p_1)$ and the
Gale--Ryser criterion. This improvement is also not accounted for in
the bounds above.

{Algorithm~\ref{algc}} traverses a directed acyclic graph
in which each node represents a distinct set of input arguments
$(\mathbf{p},\bq
)$ to the algorithm.
Node $(\u,\bv)$ is the child of node $(\mathbf{p},\bq)$ if the
algorithm is
called (recursively) with
arguments $(\u,\bv)$ while executing a call with arguments $(\mathbf
{p},\bq)$.
We associate with each node its \textit{count}: the number of matrices
with the corresponding margins.
If the initial input arguments are $(\mathbf{p},\bq)$, then all nodes are
descendants of node $(\mathbf{p},\bq)$.
Meanwhile, all nodes with positive count are ancestors of node $(\0,\0)$.
Note the correspondence between the children of a node $(\u,\bv)$ and the
compositions $\s\in C^{\bv}(u_1)$ in the binary case, and $\s\in
C^{\bv
+L\s}(u_1)$ in the $\N$-valued case, under which $\s$ corresponds with
the child $(L\u,\bv\reduce\s)$.

Once the counting is complete, these counts yield an efficient algorithm
for uniform sampling from the set of $(\mathbf{p},\mathbf{q})$ matrices
(binary or $\N$-valued).
It is straightforward to see that since the counts are exact,
the following algorithm yields a sample from the uniform distribution.

%
\begin{algorithm}[(Sampling)] \label{algs} \\
Input:
\\ $\bullet$ Row and column sums $(\mathbf{p},\mathbf{q})\in\N^m\x
\N^n$ such
that $\sum_i
p_i=\sum_i q_i$.
\\ $\bullet$ Lookup table of counts generated by {Algorithm~\ref{algc}}
on input $(\mathbf{p},\bq)$. \\
Output: A uniformly-drawn binary (or $\N$-valued) matrix with margins
$(\mathbf{p},\mathbf{q})$.

\begin{enumerate}
\item[(1)] Initialize $(\u,\v) \leftarrow(\mathbf{p},\mathbf{q})$.
\item[(2)] If $(\u,\v)=(\0,\0)$, exit.
\item[(3)]\label{choose} Choose a child $(L\u,\bv\reduce\s)$ of $(\u
,\bv)$
with probability proportional to its count times the number of
corresponding rows [i.e., the number of rows $\mathbf{r}\in C^\v
(u_1)$ such that
$\overline{\v-\mathbf{r}}=\bv\reduce\s$].
\item[(4)]\label{step} Choose a row $\mathbf{r}$ uniformly among the
corresponding rows.
\item[(5)]$(\u,\v) \leftarrow(L\u,\v-\mathbf{r})$.
\item[(6)] Go to \textup{(2)}.
\end{enumerate}
\end{algorithm}

In step (3), there are ${\bv\choose\s}$ corresponding rows
$\mathbf{r}$ in the binary case (in which $\mathbf{r}\in\{0,1\}^n$),
and ${\bv+L\s\choose\s}$ in the $\N$-valued case.
In {Section~\ref{sectioncomputation}}, we prove that {Algorithm~\ref{algs}}
takes $\O(mc\log c)$ expected time per sample, where $c=\sum_i p_i$.

A software implementation of the algorithms above has been made available,
and contains demonstrations of the applications in {Section~\ref{sectionmotivating-example}} above and {Section~\ref{sectionapplications}} below.

\section{Applications}
\label{sectionapplications}

\subsection{Zero--one tables}
\label{subsectionecology}

In several fields of study, including neurophysiology (multivariate
binary time series), sociology (affiliation matrices), psychometrics\vadjust{\goodbreak}
(item response theory) and ecology (co-occurrence matrices), the
uniform distribution on binary matrices with fixed margins is
applicable, particularly in the context of conditional inference.
In this section, we take a closer look at the ecology application,
continuing the discussion from {Section~\ref{sectionmotivating-example}}. It turns out that many
real-world ecology datasets can be accommodated by our method: out of
291 ecology matrices in a collection compiled by \citet{Atmar1995}, it
could handle~225.

In {Section~\ref{sectionmotivating-example}}, we compared our
approach to a heuristic approximation. However, it is now more common
for researchers to use MCMC [e.g., \citet{Gotelli2002,Ulrich2007}],
and recently, sequential importance sampling (SIS) approaches have been
developed that appear to improve upon MCMC [\citet{Chen2005,Harrison2013}].

%
\begin{table}
\tabcolsep=0pt
\caption{13 species of finch in 17 of the Gal\'{a}pagos Islands
[Chen et~al. (\citeyear{Chen2005})]}\label{tableDarwin}
\begin{tabular*}{\textwidth}{@{\extracolsep{\fill}}lccccccccccccccccc@{}}
\hline
& \multicolumn{17}{c@{}}{\textbf{Habitat}} \\[-6pt]
& \multicolumn{17}{c@{}}{\hrulefill} \\
\multicolumn{1}{@{}l}{\textbf{Species}} & \textbf{01} & \textbf{02}
& \textbf{03} & \textbf{04} & \textbf{05} & \textbf{06} & \textbf
{07} & \textbf{08} &
\textbf{09} & \textbf{10} & \textbf{11} & \textbf{12} & \textbf
{13} & \textbf{14} & \textbf{15} & \textbf{16} & \textbf{17} \\
\hline
A & 0 & 0 & 1 & 1 & 1 & 1 & 1 & 1 & 1 & 1 & 0 & 1 & 1 & 1 & 1 & 1 & 1 \\
B & 1 & 1 & 1 & 1 & 1 & 1 & 1 & 1 & 1 & 1 & 0 & 1 & 0 & 1 & 1 & 0 & 0 \\
C & 1 & 1 & 1 & 1 & 1 & 1 & 1 & 1 & 1 & 1 & 1 & 1 & 0 & 1 & 1 & 0 & 0 \\
D & 0 & 0 & 1 & 1 & 1 & 0 & 0 & 1 & 0 & 1 & 0 & 1 & 1 & 0 & 1 & 1 & 1 \\
E & 1 & 1 & 1 & 0 & 1 & 1 & 1 & 1 & 1 & 1 & 0 & 1 & 0 & 1 & 1 & 0 & 0 \\
F & 0 & 0 & 0 & 0 & 0 & 0 & 0 & 0 & 0 & 0 & 1 & 0 & 1 & 0 & 0 & 0 & 0 \\
G & 0 & 0 & 1 & 1 & 1 & 1 & 1 & 1 & 1 & 0 & 0 & 1 & 0 & 1 & 1 & 0 & 0 \\
H & 0 & 0 & 0 & 0 & 0 & 0 & 0 & 0 & 0 & 0 & 0 & 1 & 0 & 0 & 0 & 0 & 0 \\
I & 0 & 0 & 1 & 1 & 1 & 1 & 1 & 1 & 1 & 1 & 0 & 1 & 0 & 0 & 1 & 0 & 0 \\
J & 0 & 0 & 1 & 1 & 1 & 1 & 1 & 1 & 1 & 1 & 0 & 1 & 0 & 1 & 1 & 0 & 0 \\
K & 0 & 0 & 1 & 1 & 1 & 0 & 1 & 1 & 0 & 1 & 0 & 0 & 0 & 0 & 0 & 0 & 0 \\
L & 0 & 0 & 1 & 1 & 0 & 0 & 0 & 0 & 0 & 0 & 0 & 0 & 0 & 0 & 0 & 0 & 0 \\
M & 1 & 1 & 1 & 1 & 1 & 1 & 1 & 1 & 1 & 1 & 1 & 1 & 1 & 1 & 1 & 1 & 1 \\
\hline
\end{tabular*}
\end{table}

%

To compare with these alternatives, we consider a benchmark dataset for
this application.
{Table~\ref{tableDarwin}} indicates the presence/absence of 13
species of finch on 17 of the Gal\'{a}pagos Islands. It comes equipped
with the colorful name ``Darwin's finches'' because Charles Darwin's
development of the theory of evolution was inspired in part by his
observations of these birds.
The row and column sums of the matrix are \mbox{(14, 13, 14, 10, 12, 2, 10,
1, 10, 11, 6, 2, 17)} and (4, 4, 11, 10, 10, 8, 9, 10, 8, 9, 3, 10, 4,
7, 9, 3, 3), respectively. This table is the subject of an analysis by
Chen et al. (\citeyear{Chen2005}), in which they use MCMC and their
SIS algorithm to estimate the $p$-value for the $\overline{S^2}$
statistic of \citet{Roberts1990},
\[
\overline{S^2} = \frac{1}{{m\choose2}}\sum_{i<j}
s_{ij}^2,
\]
where $\boldS= (s_{ij}) =\A\A^{\transpose}$ and $\A$ is an
$m\times n$
co-occurrence matrix.
For given margins, larger values of $\overline{S^2}$ are interpreted as
indicating greater competition and/or cooperation among species; see
the \hyperref[appendix]{Appendix} for details.

%
\begin{table}
\caption{Sample statistics of $\overline{S^2}$ for Darwin's finch data
({Table~\protect\ref{tableDarwin}})}\label{tableDarwin-results}
\begin{tabular*}{300pt}{@{\extracolsep{\fill}}lcc@{}}
\hline
\textbf{Method} & \textbf{\# samples} & \multicolumn
{1}{c@{}}{\textbf{Estimated $\bolds{p}$-value}} \\
\hline
Exact & \phantom{0}1,000,000 & $(4.67\pm0.22)\times10^{-4} $ \\
SIS & \phantom{0}1,000,000 & $(3.96\pm0.36)\times10^{-4}$ \\
MCMC & 15,000,000 & $(3.56\pm0.68)\times10^{-4}$ \\
\hline
\end{tabular*}
\end{table}

The results of Chen et al. are reported in {Table~\ref{tableDarwin-results}}, alongside our results using exact
sampling. Our results largely confirm the conclusion of Chen et
al.---the $p$-value
is small, leading one to reject the null hypothesis;
see {Figure~\ref{figureDarwin}}.
In the \hyperref[appendix]{Appendix}, we examine the sensitivity of this
$p$-value to possible data-collection errors and to the choice of test
statistic.

%
\begin{figure}[b]

\includegraphics{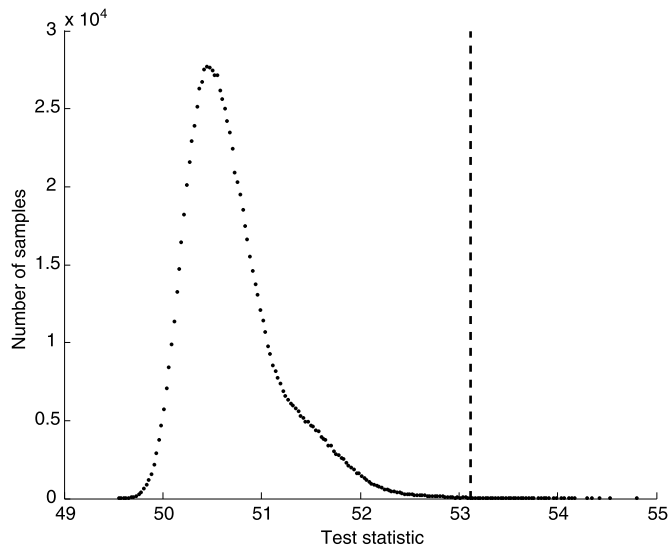}

\caption{Histogram of $\overline{S^2}$ for $10^6$ exact samples from
the uniform distribution over matrices with
margins as in {Table~\protect\ref{tableDarwin}}. The dashed line
is the value of $\overline{S^2}$ for {Table~\protect\ref
{tableDarwin}}.}\label{figureDarwin}
\end{figure}

The computation time for our method is very competitive: for this
dataset, counting the number of matrices takes 0.02 seconds, and exact
sampling takes 0.16~milliseconds/sample, compared to 1.1
milliseconds/sample for SIS and 0.07~milliseconds/sample for MCMC.
(However, on large matrices, MCMC and SIS should be significantly
faster than our method, due to the overhead incurred by counting.) The
SIS algorithm of Chen et al. also yields an estimate of $6.7150\times
10^{16}$ for the number of matrices. We compute the exact number of
matrices to be 67,149,106,137,567,626, and note that this agrees with
the exact number reported by Chen et al. (which they obtained by other means).

We should emphasize, however, that in comparison with MCMC and SIS, the
appeal of our method is not its speed, but rather its exactness. There
are no guarantees that the estimates coming from MCMC or SIS have
adequately converged. Meanwhile, we are drawing exact i.i.d. samples
from the null distribution, which provides many guarantees. For
example, using the $10^6$ exact samples above, we obtain an exact $95\%
$ confidence interval of $[4.26,5.12]\times10^{-4}$ for the $p$-value.
[By ``exact,'' we mean that it is a true confidence interval, with no
approximations involved. Reported confidence intervals often involve
two approximations: (1) approximate samples used for estimation, and/or
(2) approximate interval construction based on asymptotics, such as the
normal approximation to the binomial.]
In a longer run of $10^9$ exact samples, we estimate the $p$-value to
be $4.672\times10^{-4}$ and obtain an exact $95\%$ confidence interval
of $[4.66,4.69]\times10^{-4}$ (or $99.9\%$ interval of
$[4.65,4.70]\times10^{-4}$).
Interestingly, this only barely overlaps with the SIS approximate $95\%
$ confidence interval, $[3.25,4.67]\times10^{-4}$. Clearly, it is
advantageous to use the exact method.

\subsection{Contingency tables}
\label{subsectioncontingency-tables}

Two-way contingency tables are a large class of $\N$-valued matrices
that arise frequently in statistics. Pearson's chi-square is the
classical test of independence in such tables, however, when the
independence hypothesis fails, it can be misleading to interpret the
chi-square $p$-value as a measure of the deviation from independence.
As a starting point for quantifying the departure from independence,
\citet{Diaconis1985} propose the conditional volume test (CVT): place
the uniform distribution over tables with the observed margins, and
compute the probability of observing a chi-square value less than that
of the observed table (in other words, compute the $p$-value under this
new null distribution). Larger values of the CVT $p$-value indicate
greater deviation from independence.

Since our method enables exact sampling from the CVT null distribution,
it can be used to estimate the $p$-value for the CVT. While many
contingency tables that arise in practice will have margins that are
too large for our method, some will fall within the feasible range.

%
\begin{table}
\caption{Heights of 205 married couples [Galton (\citeyear{Galton1889})], and two related tables}\label{tableGalton}
%
\begin{tabular*}{\textwidth}{@{\extracolsep{\fill}}l@{\hspace*{-10pt}}ccc@{}}
\begin{tabular*}{10pt}{@{\extracolsep{\fill}}l@{}}
\phantom{}\\[2pt]
t \\
m\\
s\\
\phantom{}
\end{tabular*}
&
\begin{tabular*}{100pt}{@{\extracolsep{\fill}}|lcc|@{}}
\multicolumn{3}{c}{(a)}\\
\hline\noalign{\vspace*{-4pt}}
12 & 20 & 18 \\
25 & 51 & 28 \\
\phantom{0}9 & 28 & 14 \\[-3pt]
\hline\noalign{\vspace*{-4pt}}
\multicolumn{1}{c}{s}&\multicolumn{1}{c}{m}&\multicolumn{1}{c}{t}\\
\end{tabular*}
&
\begin{tabular*}{100pt}{@{\extracolsep{\fill}}|lcc|@{}}
\multicolumn{3}{c}{(b)}\\
\hline\noalign{\vspace*{-4pt}}
\phantom{0}8 & 14 & 28 \\
20 & 61 & 23 \\
18 & 24 & \phantom{0}9 \\[-3pt]
\hline\noalign{\vspace*{-4pt}}
\multicolumn{1}{c}{}&\multicolumn{1}{c}{}&\multicolumn{1}{c}{}\\
\end{tabular*}
&
\begin{tabular*}{100pt}{@{\extracolsep{\fill}}|lcc|@{}}
\multicolumn{3}{c}{(c)}\\
\hline\noalign{\vspace*{-4pt}}
16 & \phantom{0}28 & 56 \\
40 & 122 & 46 \\
36 & \phantom{0}48 & 18 \\[-3pt]
\hline\noalign{\vspace*{-4pt}}
\multicolumn{1}{c}{}&\multicolumn{1}{c}{}&\multicolumn{1}{c}{}\\
\end{tabular*}
\end{tabular*}\vspace*{3pt}
\end{table}

%
\begin{table}[b]\vspace*{3pt}
\caption{Pearson's chi-square and the conditional volume test of
Diaconis and Efron (\citeyear{Diaconis1985})}\label{tableGalton-results}
\begin{tabular*}{\textwidth}{@{\extracolsep{\fill}}lccc@{}}
\hline
& \textbf{Chi-square $\bolds{p}$-value} & \textbf{CVT $\bolds{p}$-value} & \textbf
{CVT $\bolds{95\%}$ CI} \\
\hline
{Table~\ref{tableGalton}}(a) & $0.57$ & $0.0011$ & $[0.0005,
0.0020]$ \\
{Table~\ref{tableGalton}}(b) & $1.2\times10^{-5}$ & $0.13$ &
$[0.121, 0.136]$ \\
{Table~\ref{tableGalton}}(c) & $1.8\times10^{-11}$ & $0.13$ &
$[0.123, 0.137]$ \\
\hline
\end{tabular*}
\end{table}

To illustrate, {Table~\ref{tableGalton}}(a) shows Francis
Galton's (\citeyear{Galton1889}) data recording the heights (s${} = {}$short, m${} = {}$medium, t${} = {}$tall) of $205$ married couples (e.g.,
medium--short, tall--tall, etc.). {Table~\ref{tableGalton}}(b) has
the same margins but different entries, and in {Table~\ref{tableGalton}}(c), every entry is exactly double that of
{Table~\ref{tableGalton}}(b).
For (a), (b) and (c), {Table~\ref{tableGalton-results}} shows the
results of Pearson's chi-square test\ (the approximate $p$-value) and
the conditional volume test (the estimated $p$-value and exact $95\%$
confidence interval).

The chi-square test and the CVT both indicate that (a) is close to
independence, and that (b) is not close to independence. For (c), a na\"
{i}ve interpretation of the chi-square $p$-value would be that (c) is
far further from independence than (b); meanwhile, the CVT indicates
that it deviates from independence by roughly the same amount as (b),
as one would expect.

These CVT results were each obtained using $10^4$ exact samples, drawn
at a rate of $0.14$ seconds/sample for {Tables~\ref{tableGalton}}(a)
and (b) (after $0.3$ seconds required to
count the 1,268,792 tables), and $1.94$ seconds/sample for {Table~\ref{tableGalton}}(c) (after $7.7$ seconds required to count the
19,151,218 tables).

We test our method further with two examples from \citet
{Diaconis1995}. The first is an artificial $5\times3$ table, with
margins $(10,62,13,11,39)$, $(65,25,45)$. Our algorithm takes $2.3$
seconds to count the 239,382,173 corresponding tables, and $0.3$ seconds/sample.
Their second example is significantly more challenging: a $4\times4$
table recording the eye color and hair color of 592 subjects, with
margins $(220,215,93,64)$, $(108,286,71,127)$. Our algorithm takes
16,145 seconds to count the 1,225,914,276,768,514 corresponding tables,
and $90$ seconds/sample.
These counting results match the exact numbers reported by \citet
{Diaconis1995}, obtained by other means.\looseness=-1

Our method is not well-suited to small tables with large margins (like
the last example), since it exploits the symmetries that arise when
there are many columns. For small tables, there are significantly
faster algorithms for counting, such as LattE [\citet{Deloera2004}].
In particular, these algorithms can handle extremely large margins
(while ours cannot).
However, such algorithms do not scale well to larger tables. In
contrast, our method can handle somewhat larger tables, as long as the
margins are sufficiently small---consider, for example, the
$100\times100$ example at the end of {Section~\ref{sectionoverview}}.\vadjust{\goodbreak}

\section{Proof of recursions}
\label{sectionproofs}

In this section, we prove {Theorem~\ref{thmrec}},
making rigorous the intuitive argument given in {Section~\ref{sectionidea}}.
As an alternative to the ``direct'' proof below, in \citet
{Miller2011} we also provide a generating function proof that employs
some of the beautiful properties of symmetric functions
and yields results of a more general nature.

For $\mathbf{r},\s\in\N^\infty$, let $\mathbf{r}\wedge\s$
denote the component-wise
minimum, that is, $(r_1\wedge s_1,r_2\wedge s_2,\ldots)$. In
particular, $\mathbf{r}\wedge\1 =(r_1\wedge1,r_2\wedge1,\ldots)$.
Recall our convention that $\N^n$ is considered to be the subset of
$\N^\infty$ such that all but the first $n$ components are zero.\

\begin{pf*}{Proof of {Theorem~\ref{thmrec}}}
(1) Consider the binary case. Let $(\mathbf{p},\mathbf{q})\in\N^m\x
\N^n$.
If $\mathbf{r}=\bq$, then
\begin{eqnarray*}
\bN(\mathbf{p},\mathbf{r})=N(\mathbf{p},\mathbf{q}) &\stackrel {\mathrm{(a)}}
{=}&
\sum_{\u\in C^{\mathbf{q}
\wedge\1}(p_1)} N(L\mathbf{p},\mathbf{q}-\u)
\\
&\stackrel{\mathrm{(b)}} {=}& \mathop{\sum_{\s\in\N^\infty\dvtx
}}_{\sum s_i = p_1}
\mathop{ \sum_{\u\in C^{\mathbf{q}\wedge\1}(p_1)\dvtx}}_{\overline{\mathbf
{q}-\u} =\bq\reduce\s
} N(L\mathbf{p},
\mathbf{q}-\u)\\
& \stackrel{\mathrm{(c)}} {=}& \sum_{\s\in C^\mathbf{r}(p_1)}
\pmatrix {\mathbf{r}
\cr
\s} \bN(L\mathbf{p},\mathbf{r} \reduce\s).
\end{eqnarray*}
Step (a) follows from partitioning the set of $(\mathbf{p},\mathbf
{q})$ matrices
according to the first row $\u\in C^{\mathbf{q}\wedge\1}(p_1)$ of
the matrix.
Now, it is straightforward to check that for any $\u\in C^\mathbf{q}(p_1)$
there is a unique $\s\in\N^\infty$ such that $\overline{\mathbf
{q}-\u}
=\bq
\reduce\s$ and $\sum s_i<\infty$, namely $s_i =\sum_{j=i}^\infty
(\bar q_j-(\overline{\mathbf{q}-\u})_j )$, and it follows that $\sum s_i
= p_1$.
Step (b) partitions $C^{\mathbf{q}\wedge\1}(p_1)$ into the level sets
of the
map taking $\u$ to this $\s\in\N^\infty$.
Step (c) follows since if $\overline{\mathbf{q}-\u}=\bq\reduce\s$
and $\mathbf{r}
=\bq$,
then $N(L\mathbf{p},\mathbf{q}-\u)=\bN(L\mathbf{p},\overline
{\mathbf{q}-\u
})=\bN(L\mathbf{p},\mathbf{r}\reduce\s)$, and
by {Lemma~\ref{lemmacounts}}(1) (below) the inner sum contains
${\mathbf{r}\choose\s}$ terms. The range of the sum reduces to
$C^\mathbf{r}(p_1)$
since ${\mathbf{r}\choose\s}$ is zero unless $\mathbf{r}\leq\s$.

(2) The $\N$-valued case is nearly identical, with the obvious changes
[replace $C^{\mathbf{q}\wedge\1}(p_1)$, $C^\mathbf{r}(p_1)$ and
${\mathbf{r}\choose\s}$
by $C^\mathbf{q}
(p_1)$, $C^{\mathbf{r}+ L\s}(p_1)$ and ${\mathbf{r}+ L\s\choose\s
}$, resp.,
and use {Lemma~\ref{lemmacounts}}(2)].
\end{pf*}

%
\begin{lemma}
\label{lemmacounts}
If $\v\in\N^n$, $\s\in\N^\infty$ and $k =\sum s_i<\infty$, then:
\begin{longlist}[{(1)}]
\item[{(1)}] $\ds\# \{\u\in C^{\v\wedge\1}(k) \dvtx
\overline{\v
-\u} =\bar\v\reduce\s\} ={\bar\v\choose\s}$ and
\item[{(2)}] $\ds\# \{\u\in C^{\v}(k) \dvtx\overline{\v
-\u} =\bar
\v\reduce\s\} ={\bar\v+ L\s\choose\s}$.
\end{longlist}
\end{lemma}
\begin{pf}
(1) Let $b =\max v_i$. If $\u\in C^{\v\wedge\1}(k)$ and $\overline
{\v-\u
} =\bar\v\reduce\s$, then $s_i = 0$ for $i>b$, and for $i = b$,
$b-1,\ldots,1$ (starting with $i = b$ and working our way down), the
fact that $(\overline{\v-\u})_i =\bar v_i-s_i + s_{i+1}$ implies that
out of the $\bar v_i$ places $j$ such that $v_j = i$, a subset of
exactly $s_i$ of them has $u_j = 1$. On the other hand, for any such
sequence of subsets, there is a unique $\u\in C^{\v\wedge\1}(k)$
satisfying $\overline{\v-\u} =\bar\v\reduce\s$ which gives rise
to it
in this way. Clearly, there are ${\bar\v\choose\s}$ such sequences
of subsets.

(2) We may as well assume $\s\leq\bar\v+ L\s$, since otherwise both
sides are trivially zero. Let $b =\max v_i$, and note that $s_i = 0$
for all $i>b$. Consider the following algorithm for choosing a $\u\in
C^\v(k)$ such that $\overline{\v-\u} =\bar\v\reduce\s$.
Initialize $\u$
to be identically zero. There are $\bar v_b$ places $j$ where $v_j =
b$, so in order to satisfy $(\overline{\v-\u})_b =\bar v_b-s_b$,
increment $u_j\leftarrow u_j +1$ for $s_b$ of these places (and keep
the other $\bar v_b-s_b$ unchanged). Then after this, there are $\bar
v_{b -1} + s_b$ places $j$ where $v_j-u_j = b -1$ (namely, the $\bar
v_{b -1}$ places where $v_j = b -1$, along with the $s_b$ places where
$v_j = b$ and $u_j = 1$ from the previous step), so to satisfy
$(\overline{\v-\u})_{b -1} =\bar v_{b -1}-s_{b -1} + s_b$, increment
$u_j\leftarrow u_j +1$ for $s_{b -1}$ of these places. Continuing in
this way, for $i = b -2,\ldots,1$: there are $\bar v_i + s_{i +1}$
places $j$ where $v_j-u_j = i$ (the $\bar v_i$ places where $v_j = i$,
along with the $s_{i+1}$ places where $v_j>i$ and $u_j = v_j-i$ due to
previous steps), so to satisfy $(\overline{\v-\u})_i =\bar v_i-s_i +
s_{i+1}$, increment $u_j\leftarrow u_j +1$ for $s_i$ of these places.

Clearly, there are
\[
\pmatrix{\bar v_b
\cr
s_b} \pmatrix{\bar v_{b -1}
+ s_b
\cr
s_{b -1}}\cdots\pmatrix{\bar v_1+
s_2
\cr
s_1} =\pmatrix{\bar\v+ L\s
\cr
\s}
\]
ways to follow this algorithm. By construction, any $\u$ obtained via
the algorithm satisfies $(\overline{\v-\u})_i =\bar v_i-s_i + s_{i+1}$
for each $i= 1,\ldots,b$, $ \0\leq\u\leq\v$ and $\sum_{j = 1}^n
u_j =
\sum_{i = 1}^b s_i$, and therefore $\u\in C^\v(k)$ and $\overline
{\v-\u
} =\bar\v\reduce\s$. Meanwhile, given any $\u\in C^\v(k)$ such that
$\overline{\v-\u} =\bar\v\reduce\s$, there is a unique way to
obtain it
via the algorithm, since for $i = b,\ldots,1$, the choice of which
$\bar v_i-s_i + s_{i+1}$ places to leave unchanged is uniquely
determined. Therefore, the number of ways to follow the algorithm
equals the number of $\u\in C^\v(k)$ such that $\overline{\v-\u}
=\bar\v
\reduce\s$.
\end{pf}

\section{Computation time}
\label{sectioncomputation}

Let $W(\mathbf{r}):=\sum_{k=1}^n kr_k=$ the \textit{weight} of
$\mathbf{r}\in\Z^n$.

%
\begin{lemma}[(Properties of the weight)] If $\mathbf{r},\s\in\Z^n$, then:
\label{lemW}
\begin{longlist}[(1)]
\item[(1)]$W(\mathbf{r}+\s)=W(\mathbf{r})+W(\s)$;
\item[(2)]$W(\s-L\s)=\sum s_i$;
\item[(3)]$W(\mathbf{r}\reduce\s)=W(\mathbf{r})-\sum s_i$;
\item[(4)]$W(\bar\s) = \sum s_i$.
\end{longlist}
\end{lemma}
\begin{pf}
All four are simple calculations.
\end{pf}

For the rest of this section, fix $(\mathbf{p},\mathbf{q})\in\N^m\x
\N^n$
such that $\sum p_i=\sum q_i$, and
consider $(\mathbf{p},\mathbf{q})$ to be the margins of a set of $m\x
n$ matrices.
First, we address the time to compute $N(\mathbf{p},\mathbf{q})$ using
{Algorithm~\ref{algc}},
and $M(\mathbf{p},\mathbf{q})$ will follow easily.

Let $\D(\mathbf{p},\mathbf{q})$ denote the set of nontrivial nodes
$(\u,\bv
)$ in the
directed acyclic graph
(as discussed in {Section~\ref{sectionresults}})
descending from $(\mathbf{p},\bq)$ [including $(\mathbf{p},\bar
\mathbf{q}
)$], where nontrivial
means $(\u,\bv)\neq(\0,\0)$.
Let $\Delta_k(j):=\{\s\in\N^k \dvtx W(\s)=j\}$ for $j,k\in\N$.
The intuitive content of the following lemma is that the graph
descending from $(\mathbf{p},\bq)$ is
contained in a union of sets $\Delta_k(j)$ with weights decreasing by
steps of
$p_1,\ldots,p_m$.

%
\begin{lemma}[(Descendants)]
\label{lemL}
If $t_j =\sum_{i=j}^m p_i$ and $b =\max q_i$, then
\[
\D(\mathbf{p},\mathbf{q}) \sbs \bigl\{(\u,\bv) \dvtx\u =L^{j-1}
\mathbf{p}, \bv \in\Delta_b(t_j), j=1,\ldots,m \bigr\}.
\]
\end{lemma}
\begin{pf}
By the form of the recursion, $(\u,\bv)\in\D(\mathbf{p},\mathbf
{q})$ if
and only if for
some $1\le j\le m$ there exist
$\s^1,\ldots,\s^{j-1}$ in $C^{\mathbf{r}^1}(p_1),\ldots,C^{\mathbf
{r}^{j-1}}(p_{j-1})$,
respectively, with
$\mathbf{r}^1=\bq$, $\mathbf{r}^{i+1}=\mathbf{r}^i\reduce\s^i$
for $i=1,\ldots,j-1$,
such that
$(\u,\bv)=(L^{j-1}\mathbf{p},\mathbf{r}^j)$.
For $j\geq2$, by {Lemma~\ref{lemW}} (3 and 4),
\begin{eqnarray*}
W \bigl(\mathbf{r}^j \bigr)&=&W \bigl(\mathbf{r}^{j-1}\reduce
\s^{j-1} \bigr)=W \bigl(\mathbf{r}^{j-1} \bigr)-p_{j-1}=W
\bigl(\mathbf{r}^{j-2} \bigr)-p_{j-2}-p_{j-1}
\\
&=&\cdots=W \bigl(\mathbf{r}^1 \bigr)-(p_1+
\cdots+p_{j-1})\\
&=& \sum_{i = 1}^n
q_i-\sum_{i =
1}^{j-1}
p_i = \sum_{i = 1}^m
p_i-\sum_{i = 1}^{j-1}
p_i = t_j
\end{eqnarray*}
and $\mathbf{r}^j\in\N^b$ by construction, so
$\mathbf{r}^j\in\Delta_b(t_j)$.
Hence, $(\u,\bv)=(L^{j-1}\mathbf{p},\mathbf{r}^j)$ belongs to the
set as claimed.
\end{pf}

Let $T(\mathbf{p},\mathbf{q})$ be the time (number of machine operations)
required by
the algorithm
({Algorithm~\ref{algc}}) to compute $N(\mathbf{p},\mathbf{q})$ after
precomputing
all needed binomial coefficients.
Let $\t(\u,\bv)$ be the time to compute $\bN(\u,\bv)$ given
$\bN(L\u,\bv\reduce\s)$ for all $\s\in C^\bv(u_1)$. That is,
$T(\mathbf{p},\mathbf{q})$
is the
time to perform the entire recursive computation, whereas $\t(\u,\bv)$
is the time to perform
a given call to the algorithm not including time spent in subcalls.

Let $n_0:=\#\{i\dvtx q_i>0\}$ denote the number of nonempty columns.
By constructing Pascal's triangle, we precompute all of the binomial
coefficients that may be needed, and store them in a lookup table. We
only need binomial coefficients with entries less or equal to $n_0$,
for the following reason. In the binary case, the recursion involves
numbers of the form ${\bar\v\choose\s}$ with $\s\leq\bar\v$,
and for
any descendent $(\u,\bar\v)$ and any $i>0$ we have $\bar v_i\leq n_0$
since the number of columns with sum $i$ is less or equal to the total
number of nonempty columns. For the $\N$-valued case, the same set of
binomial coefficients will be sufficient, since then we have numbers of
the form ${\bar\v+ L\s\choose\s}$ with $\s\leq\bar\v+ L\s$,
and thus
\[
\bar v_i + s_{i+1}\leq\bar v_i + \bar
v_{i+1} + s_{i+2}\leq\cdots\leq\bar v_i + \bar
v_{i+1} +\bar v_{i+2} + \cdots\leq n_0,
\]
where the last inequality holds because the number of columns $j$ with
sum greater or equal to $i$ is no more than the total number of
nonempty columns. Since the addition of two $d$-digit numbers takes
$\Theta(d)$ time, and there are ${n_0 +2\choose2}$ binomial
coefficients with entries less or equal to $n_0$, then the bound $\log
_{10} {j \choose k} +1 \leq n_0\log_{10} 2 +1 $ on the number of digits
for such a binomial coefficient shows that this pre-computation can be
done in $\O(n_0^3)$ time. Except in trivial cases (when the largest
column sum is 1), the additional time needed does not affect the bounds
on $T(\mathbf{p},\mathbf{q})$ that we will prove below.

We now bound the time required for a given call to the algorithm.

%
\begin{lemma}[(Time per call)]\label{lemt} $\t(\u,\bv)=\O
((ab+c)(\log c)^3
|C_b(u_1)|)$ for $(\u,\bv)\in\D(\mathbf{p},\mathbf{q})$,
where $a =\max p_i, b =\max q_i$ and $c =\sum p_i$.
\end{lemma}
\begin{pf}
See \citet{Miller2011}. Due to space constraints, the proof has been omitted.
\end{pf}

%
\begin{lemma}\label{lemD} $\ds\#\Delta_k(j)\leq{j+k-1 \choose k-1}$
for any
$j,k\in\N$.
\end{lemma}
\begin{pf}
The map $f(\mathbf{r})=(1r_1,2r_2,\ldots,kr_k)$ is an injection
$f\dvtx\Delta
_k(j)\to C_k(j)$.
Thus $\#\Delta_k(j)\leq\#C_k(j)={j+k-1 \choose k-1}$.
\end{pf}

We are now ready to prove {Theorem~\ref{thmeff}}.

\begin{pf*}{Proof of Theorem~\ref{thmeff} for
$N(\mathbf{p},\mathbf{q})$}
By storing intermediate results in a lookup table,
once we have computed $\bN(\u,\bv)$ upon our first visit to
node $(\u,\bv)$, we can simply reuse the result for later visits.
Hence, we need only expend $\t(\u,\bv)$ time for each node $(\u,\bv
)$ occuring
in the graph. Let $t_j =\sum_{i=j}^m p_i$ and $d=(ab+c)(\log c)^3$. Then
\begin{eqnarray*}
T(\mathbf{p},\mathbf{q}) = \sum_{(\u,\bv)\in\D(\mathbf
{p},\mathbf{q})} \!\!\t(\u,\bv)
&\stackrel{\mathrm{(a)}} {\leq}& \sum_{j=1}^m
\sum_{\bv\in\Delta_b(t_j)} \t \bigl(L^{j-1}\mathbf{p},\bv
\bigr)
\\
&\stackrel{\mathrm{(b)}} {\leq}& \sum_j \sum
_\bv\O \bigl(d\bigl |C_b(p_j)\bigr|
\bigr) = \sum_j \O \bigl(d\bigl|C_b(p_j)\bigr|
\bigl| \Delta_b(t_j)\bigr| \bigr)
\\
&\stackrel{\mathrm{(c)}} {\leq}& \sum_j \O\biggl(d
\pmatrix{p_j+b-1
\cr
b-1} \pmatrix{t_j+b-1
\cr
b-1}\biggr)
\\
&\stackrel{\mathrm{(d)}} {\leq}& \sum_j \O\biggl(d
\pmatrix{a+b-1
\cr
b-1} \pmatrix{c+b-1
\cr
b-1}\biggr)
\\
&\leq&\O \bigl(dm (a+b-1)^{b-1}(c+b-1)^{b-1} \bigr),
\end{eqnarray*}
where (a) follows by {Lemma~\ref{lemL}}, (b) by {Lemma~\ref{lemt}},
(c) by {Lemma~\ref{lemD}} and
(d) since $p_j\leq a$ and $t_j\leq c$.
This proves (1) and (2).
Now, (3) and (4) follow from (2) since $a \leq c \leq bn$.
\end{pf*}

\begin{pf*}{Proof of {Theorem~\ref{thmeff}} for
$M(\mathbf{p},\mathbf{q})$}
Other than the coefficients, the only difference between the recursion
for $\bM(\mathbf{p},\bq)$ and
that for $\bN(\mathbf{p},\bq)$ is that we are summing over $\s$
such that $\s
\in C^{\mathbf{r}+L\s}(p_1)$.
{Lemma~\ref{lemL}} holds with the same proof, except with
$C^{\mathbf{r}^1}(p_1),\ldots, C^{\mathbf{r}^{j-1}}(p_{j-1})$
replaced by $C^{\mathbf{r}^1+L\s^1}(p_1), \ldots,\break C^{\mathbf
{r}^{j-1}+L\s^{j-1}}(p_{j-1})$, respectively.
{Lemma~\ref{lemt}} also continues to hold; see \citet
{Miller2011} for details.
Consequently, the proof of the bounds goes through as well.
\end{pf*}

This completes the proof of {Theorem~\ref{thmeff}}.
Now we address the time required to uniformly sample a matrix with
specified margins.
Let $T_r(k)$ be the maximum over $1\leq j\leq k$ of the expected time
to generate a random integer uniformly between $1$ and $j$.
If we are given a random bitstream [independent and identically
distributed Bernoulli$(1/2)$ random variables] with constant cost per
bit, then $T_r(k) =\O(\log k)$, since for any $j \leq k$, $\lceil\log
_2 j \rceil\leq\lceil\log_2 k \rceil$ random bits can be used to
generate an integer uniformly between $1$ and $2^{\lceil\log_2
j\rceil
}$, and then rejection sampling can be used to generate uniform samples
over $\{1,\ldots,j\}$. Since the expected value of a Geometric$(p)$
random variable is $1/p$, then the expected number of samples required
to obtain one that falls in $\{1,\ldots,j\}$ is always less than $2$.
More generally, for any fixed $d\in\N$, if we can draw uniform samples
from $\{1,\ldots,d\}$, then we have $T_r(k) =\O(\log k)$ by
considering the base-$d$ analogue of the preceding argument.

%
\begin{lemma}[(Sampling time)]
\label{lemstime}
{Algorithm~\ref{algs}} takes
\[
\O \bigl(mT_r \bigl(n^c \bigr) + maT_r(n)
+ mb\log(a+b) \bigr)
\]
expected time per sample in the binary case, and
\[
\O \bigl(mT_r \bigl((2c)^c \bigr)+maT_r(n)+mb
\log(a+b) \bigr)
\]
expected time per sample in the $\N$-valued case.
If $T_r(k) = \O(\log k)$, then this is $\O(mc\log c)$ expected time per
sample in both cases.
\end{lemma}
%
%
\begin{remark*} If $b$ is bounded then $\O(mc\log c)\leq\O(mn\log n)$
since $c\leq bn$, and so this is polynomial expected time for bounded
column sums.
\end{remark*}
\begin{pf}
See \citet{Miller2011}. Due to space constraints, the proof has been omitted.
\end{pf}

\begin{appendix}
\section*{Appendix: Sensitivity analysis}
\label{appendix}

Here, we examine the sensitivity of the ecology results in
{Section~\ref{subsectionecology}} to data-collection errors and to the
choice of test statistic.
First, it is quite possible that co-occurrence matrices such as the
finch data in {Table~\ref{tableDarwin}} may contain some
data-collection errors; in particular, it is conceivable that a given
species does in fact inhabit a particular island, but was not seen by
the observers. We analyze the sensitivity of the $p$-value to such
errors, and find that it is not particularly sensitive. Second, the
particular choice of test statistic should not strongly influence the
results. We assess the sensitivity of the $p$-value to variations of
the test statistic.

\subsection{Sensitivity to data-collection errors}

\begin{figure}

\includegraphics{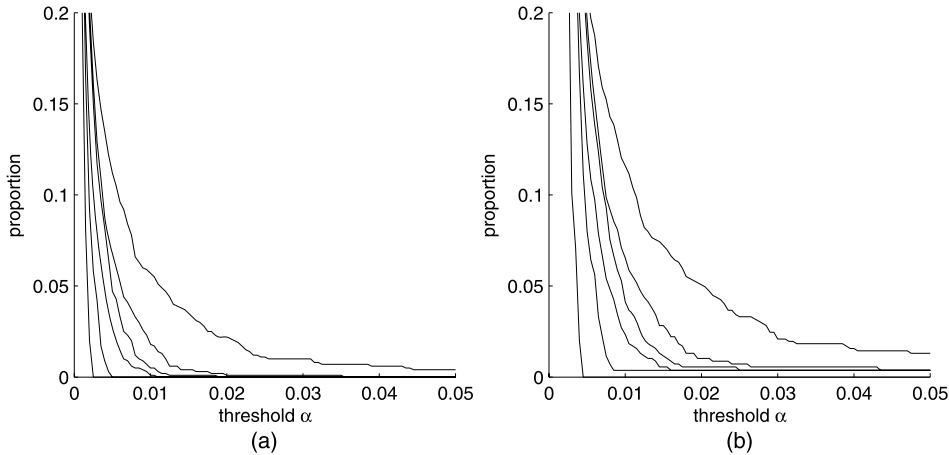}

\caption{\textup{(a)} Estimated proportion of $p$-values exceeding $\alpha$, and
\textup{(b)} $95\%$ confidence upper bound
on this proportion. From bottom to top, the curves are $d=1,2,3,4,5,10$.}
\label{figuresensitivity}
\end{figure}

Given $d\in\{1,2,\ldots\}$, consider the set of binary matrices that
can be obtained from {Table~\ref{tableDarwin}} by flipping $d$
zeros to ones. Each such matrix has a $p$-value under the test
statistic $\overline{S^2}$. For $d\in\{1,2,3,4,5,10\}$, and for a range
of threshold values $\alpha\in[0,0.05]$, we estimated the proportion of
matrices with a $p$-value exceeding $\alpha$; see {Figure~\ref{figuresensitivity}}(a). {Figure~\ref{figuresensitivity}}(b) contains
(exact) $95\%$ confidence
upper bounds on these estimates, for any given threshold $\alpha$ (not for
all thresholds simultaneously). See below for details.

%

These results suggest that the proportions are very small---indeed, the
estimated proportion exceeding $\alpha= 0.05$ is zero for all $d\leq5$.
This indicates that the $p$-value is not highly sensitive to
data-collection errors of this kind. We conducted the same analysis
over all binary matrices within Hamming distance $d$ (although it seems
unlikely that a species would be mistakenly recorded as present), and
the estimates are somewhat larger---particularly for the case of
$d=10$---but remain small for $d\leq5$.

For $d>1$, the estimates were made by uniformly drawing $L=1000$
matrices at distance $d$ and, for the $i$th matrix, using $N=10^4$
samples from our algorithm to compute an estimate $\hat\theta_i$ of the
$p$-value. For each $d$, {Figure~\ref{figuresensitivity}}(a)
shows $\alpha$ versus $L^{-1}\sum_{i=1}^L I(\hat\theta_i > \alpha)$.
Let $B_{N,\beta}(x) = \sup\{ \theta\in[0,1] \dvtx F(x;N,\theta)
\geq
\beta
\}$, where $F(x;N,\theta)=\P(X\leq x)$ with $X\sim$
Binomial$(N,\theta
)$. {Figure~\ref{figuresensitivity}}(b) shows $\alpha$ versus
$B_{L,\gamma} (\sum_{i=1}^L I(B_{N,\beta}(N\hat\theta_i) >
\alpha
) )$, for $\gamma=0.025$ and $\beta=1-(1-\gamma)^{1/L}$. The case
$d=1$ is slightly different because we exhaustively explore all $99$
possible matrices instead of sampling $L$ of them.

\subsection{Sensitivity to test statistic variations}

The idea behind the \citet{Roberts1990} test statistic $\overline{S^2}
= {m\choose2}^{-1}\sum_{i<j} s_{ij}^2$ is as follows. Uniformly choose
a pair of species, and let $X$ be the number of habitats shared [i.e.,
$X$ is $s_{ij}$ with probability $1/{m \choose2}$]. \citet
{Roberts1990} argue that, although exceptions can be constructed,
competition/cooperation effects will typically make $\Var(X)$ larger
(relative to other co-occurrence matrices with the same margins), since
they will tend to make the $s_{ij}$ values more extreme.
Using $\Var(X)$ as a test statistic is equivalent to using $\overline
{S^2}$, since $\E(X^2)=\overline{S^2}$ and $\E(X)$ is the same for all
matrices with the same margins.

It seems that the same argument could be used to justify any statistic
of the form $\E f(|X-\E X|)$, where $f$ is monotone increasing on
$[0,\infty)$. Of course, $\Var(X)$ is the case of $f(x) = x^2$. To
study sensitivity to the choice of $f$, we estimated the $p$-value of
the finch data ({Table~\ref{tableDarwin}}) with $f(x) = x^c$ for
$c\in\{0.5,1,2,3\}$, using $10^5$ samples. See {Table~\ref{tablesensitivity}}.

%
\begin{table}
\caption{Effect of the test statistic on the $p$-value for
{Table~\protect\ref{tableDarwin}}}\label{tablesensitivity}
\begin{tabular*}{300pt}{@{\extracolsep{\fill}}lcc@{}}
\hline
\multicolumn{1}{@{}l}{$\bolds{c}$} &\textbf{ Estimated $\bolds{p}$-value} & \multicolumn{1}{c@{}}{\textbf{$\bolds{95\%}$ confidence interval}} \\
\hline
$0.5$ & $0.162$\phantom{00} & $[0.159, 0.164]$ \\
$1$ & $0.0113$\phantom{0} & $[0.0106, 0.0120]$ \\
$2$ & $0.00044$ & $[0.00032, 0.00059]$ \\
$3$ & $0.00016$ & $[0.00009, 0.00026]$ \\
\hline
\end{tabular*}
\end{table}

Interestingly, the $p$-value with $c = 0.5$ is quite large. This
suggests that, in this case at least, the more extreme values of
$s_{ij}$ are playing an important role. Whether or not this is
scientifically relevant is a question for ecologists.
\end{appendix}

\section*{Acknowledgement}
We would like to thank the Associate Editor for suggesting the
sensitivity analysis in the \hyperref[appendix]{Appendix}.


%



\printaddresses

\end{document}